\documentclass[showpacs,twocolumn,preprintnumbers,amsmath,amssymb,pra]{revtex4}

\usepackage{graphicx}
\usepackage{dcolumn}
\usepackage{bm}

\renewcommand{\vec}[1]{{\mathbf #1}}

\begin{document}

\title{Bound-state signatures in quenched Bose-Einstein condensates}

\author{John P. Corson}
\author{John L. Bohn}

\affiliation{JILA, NIST and Department of Physics, University of Colorado, Boulder, Colorado 80309-0440, USA}

\date{\today}

\begin{abstract}
We investigate the dynamics of a homogenous Bose-Einstein condensate (BEC) following a sudden quench of the scattering length. Our focus is the time evolution of short-range correlations via the dynamical contact. We compute the dynamics using a combination of two- and many-body models, and we propose an intuitive connection between them that unifies their short-time, short-range predictions. Our two-body models are exactly solvable and, when properly calibrated, lead to analytic formulae for the contact dynamics. Immediately after the quench, the contact exhibits strong oscillations at the frequency of the two-body bound state. These oscillations are large in amplitude, and their time average is typically much larger than the unregularized Bogoliubov prediction. The condensate fraction shows similar oscillations, whose amplitude we are able to estimate. These results demonstrate the importance of including the bound state in descriptions of diabatically-quenched BEC experiments.
\end{abstract}

\maketitle

\section{Introduction} \label{sec:Introduction}
Recent advances in the tunability of ultracold atomic gases have opened the door for the study of interesting many-body sytems. At low energies, two-body scattering is determined by the scattering length $a$, and this can be tuned to arbitrary values near a broad Fano-Feshbach resonance \cite{Tiesinga1993,Inouye1998,Courteille1998,Vuletic1999,Chin2010}. Control in the interaction is then limited only by the stability of the external magnetic field relative to the width of the broadest available resonance.

Experimental frontiers are also expanding to consider nonequilibrium scenarios. A conceptually-simple setup for observing nonequilibrium dynamics is to quench a system parameter, such as the scattering length, and then observe the response of the system. This can be accomplished by a single, fast ramp of the magnetic field near a Fano-Feshbach resonance. Experiments have shown that a quenched Bose-Einstein condensate (BEC) can exhibit Sakharov oscillations \cite{Hung2013}, as well as nontrivial decay dynamics \cite{Claussen2002}. Most recently, Ref.~\cite{Makotyn2014} demonstrated that, if a BEC is quenched suddenly to unitarity ($a\rightarrow \infty$), the three-body loss is not as catastrophic as one would expect. This was later argued to be a consequence of the projective nature of the experiment: the initial condition projects mainly only long-lived states, thereby limiting inelastic loss \cite{Sykes2014}. Diabatic quenches thus represent a possible pathway to exotic many-body states.

The topic of quenched BECs has received considerable theoretical interest in the literature as of late. The dynamics of correlation functions at small $n a^3$ was computed recently within the Bogoliubov approximation \cite{Natu2013}. Two other studies used an unregularized, saturating effective interaction within a self-consistent Bogoliubov approximation \cite{Yin2013,Kain2014} to capture the universal density scaling observed in the unitary-Bose-gas experiment \cite{Makotyn2014}. References \cite{Rancon2013,Rancon2014} investigated the effects of phenomenological damping on the eventual equilibration of quenched BECs within the Bogoliubov approximation. Another study described BEC quench dynamics using a regularized pseudopotential and a variational many-body wave function, combined with exactly-solvable few-body models \cite{Sykes2014}. A recent quantum-kinetic-theory study has employed a short-range Morse potential in its description of quenched BECs \cite{Kira2014a,Kira2014b}.

A fact that has been underappreciated lately is that the Feshbach molecular bound state may play a dominant role in BEC quench dynamics. This state exists only on the repulsive side of a resonance ($a>0$), and its energy is $E_B=-\hbar^2/m a^2$. Bound-state physics was responsible for the pronounced Ramsey oscillations described in Refs.~\cite{Donley2002,Claussen2003}, as well as the nontrivial expansion \cite{Ronzheimer2013,Boschi2014} and spin-propagation \cite{Ganahl2012,Fukuhara2013} dynamics observed recently in one-dimensional lattice gases. We expect it to be of similar importance in diabatic quench experiments, where pairs of atoms may project nontrivially onto the post-quench bound state. Such a projection can qualitatively change the short-range correlation dynamics of the system. One way to account for this is to use a two-channel model, as was common a decade ago \cite{Timmermans1999, Holland2001, Kokkelmans2002, Mackie2002, Kohler2002, Kohler2003, Milstein2003, Duine2003a, Duine2003b, Duine2004, Goral2005, Kohler2006, Snyder2012}. Single-channel descriptions must use a regularized pseudopotential (or a short-ranged variant thereof) that admits a bound state \cite{footnote0}.

In this paper, we examine the effect of the bound state on the short-time, short-range correlations of a BEC that is quenched suddenly between two scattering lengths. We use a properly-regularized contact interaction within two- and many-body models, as done in Ref. \cite{Sykes2014}. We introduce an intuitive calibration scheme that unambiguously links the few-body models to many-body physics; our new prescription unifies the few-body predictions across a broad class of exactly-solvable models, while yielding analytic formulae that agree with less transparent, many-body numerics. Our focus is the time evolution of Tan's contact, defined by $C(t)\equiv\mathrm{lim}_{k\rightarrow\infty} k^4 n_k(t)$ for momentum distributions that are normalized to the density $n$ via $n=\int d^3k n_k(t) / (2\pi)^3$.  The contact is a measure of the short-range correlations of a system, such that the two-particle correlation function of a homogenous single-component Bose gas behaves as 
\begin{equation} \label{eq: g2}
\begin{aligned}
g^{(2)}(\vec r,t)&\equiv \left\langle \hat \psi^{\dagger}(\vec r,t) \hat{\psi}^\dagger(0,t)\hat{\psi}(0,t)\hat{\psi}(\vec r,t)\right\rangle /n^2 \\&
\rightarrow\frac{C(t)}{16\pi^2n^2 r^2} 
\end{aligned}
\end{equation}
for small $r$ \cite{Tan2008a,Tan2008b,Tan2008c}. The dynamical contact has received recent experimental attention in Ref. \cite{Bardon2014}, wherein RF pulses were used to probe the short-range correlations of a spin-diffusing Fermi gas. For the case of a BEC that undergoes a diabatic quench of the scattering length, we find that the contact exhibits strong oscillations at the frequency of the post-quench bound state, $\omega_B=\left| E_B \right|/\hbar$. If the quench is diabatic, both the oscillation amplitude and its time average may be large compared to the contact predicted by unregularized Bogoliubov theory. Strikingly, even a downward quench of the scattering length (where $a_f < a_i$) may increase the short-range correlations of the system. These interesting results are a direct consequence of bound-state physics.

Section \ref{sec: mean field} reviews our many-body variational formalism and demonstrates its relation to previous treatments of BEC near resonance. We then introduce the basic phenomenon of interest in this paper: bound-state oscillations. Next, we discuss a class of two-body models in Sec. \ref{sec: two body}, presenting a calibration scheme that unifies their predictions for short-time, short-range dynamics. We derive an analytic formula for the contact dynamics following a diabatic quench, and we describe the physical origin of the observed oscillations. Finally, Sec.~\ref{sec: conclusion} concludes our analysis.

\section{Many-body Phenomenon} \label{sec: mean field}

Although the recent experimental work described in Ref.~\cite{Makotyn2014} has drawn attention mainly to the unitary Bose gas, a similar diabatic-quench apparatus may be used to probe bound-state dynamics at finite scattering length. We describe such a quenched system using the many-body-variational approach used previously in Ref. \cite{Sykes2014}. Our focus will be the short-time, coherent dynamics of large-momentum observables.

We consider a homogenous BEC in which, after a sudden quench to scattering length $a_f$, all atoms interact via regularized contact interactions. Assuming periodic boundary conditions in a box of volume $V$, the Hamiltonian that describes such an interacting system is given by
\begin{equation} \label{eq: hamiltonian}
\hat H = \sum_{\vec k}^\Lambda\epsilon_{\vec k}\hat a_{\vec k}^\dagger \hat a_{\vec k}+\frac{U_\Lambda}{2V}  \sum_{\vec k_1, \vec k_2 ,\vec q}^\Lambda\hat a_{\vec k_1+\vec q}^\dagger \hat a_{\vec k_2-\vec q}^\dagger \hat a_{\vec k_1} \hat a_{\vec k_2} ,
\end{equation}
where $\hat a_{\vec k}$ ($\hat a_{\vec k}^\dagger$) annihilates (creates) a boson of momentum $\vec k$, $\epsilon_{\vec k} = \hbar^2k^2/2m$ is the single-particle kinetic energy, $\Lambda$ is a momentum cutoff, and 
\begin{equation} \label{eq: coupling constant}
U_\Lambda = \frac{4\pi\hbar^2a_f/m}{1-\frac{2}{\pi}\Lambda a_f}
\end{equation}
is the cutoff-dependent interaction strength. In the limit that $\Lambda \rightarrow\infty$, the interactions are truly zero range and admit a single bound state of energy $E_B~=~-\hbar^2/m a_f^2$. Although the physics of Fano-Feshbach resonances is always multi-channel in nature, entrance-channel-dominated resonances in ultracold gases are generally well approximated by short-range single-channel interactions \cite{Chin2010}, such as the one we use.

Immediately after a quench, pairs of atoms begin to scatter out of an initially-pure BEC. It thus makes sense to describe the early stages of time evolution using a variational ansatz that generates pairs of atoms from a coherent state of condensed atoms. Therefore, we use the time-dependent ansatz introduced in Ref. \cite{Sykes2014}:
\begin{equation} \label{eq: ansatz}
| \Psi(t) \rangle = \mathcal{A}(t) \mathrm{exp}\left[ c_0(t) \hat a_0^{\dagger}+\sum_{\vec k\cdot \hat{\vec z}>0}^\Lambda g_\vec k (t)\hat a_{\vec k}^{\dagger}\hat a_{-\vec k}^{\dagger}\right] |0\rangle,
\end{equation}
where $c_0(t)$ and $\left\{g_{\vec k}(t)  \right\}$ are time-dependent variational parameters, $\left|0  \right\rangle$ is the particle vacuum, and
\begin{equation} \label{eq: ansatz normalization}
\mathcal{A}(t) = \mathrm{exp}\left\{-\left|c_0(t)  \right|^2/2+\frac{1}{2}\sum_{\vec k\cdot \hat{\vec z}> 0}^\Lambda \mathrm{ln}\left[1-\left|g_{\vec k}(t)  \right|^2  \right]   \right\}
\end{equation}
is a normalization constant. It is simple to show that the variational parameters are related to the dynamic momentum populations via $n_0(t)=\left|c_0(t)  \right|^2$ and $n_{\vec k}(t) = \left|g_{\vec k}(t)  \right|^2/(1-\left|g_{\vec k}(t)  \right|^2)$. We choose to consider an initial condition representing a pure BEC of  density $n$: $c_0(0) = \sqrt{nV}$ and $g_{\vec k}(0)=0$ for all $\vec k\cdot\hat{\vec z}>0$. A time-independent version of this ansatz has been used to compute the constrained ground state of a strongly-interacting Bose gas \cite{Song2009}, with precedent from \cite{Girardeau1959,Nozieres1982}. 

We derive the equations of motion for the system by minimizing the action, where the Lagrangian is \cite{Kramer2008}
\begin{equation}
\mathcal{L} = \frac{i\hbar}{2}\left( \Big\langle \Psi(t)  \left| \dot{\Psi}(t)\right\rangle -\left\langle \dot{\Psi}(t)\right| \Psi(t)\Big\rangle \right) - \left\langle \Psi(t)  \right| \hat H \left| \Psi(t)\right\rangle .
\end{equation}
Given the ansatz \eqref{eq: ansatz}, it can be shown that the Euler-Lagrange equations of motion for the system are
\begin{equation}
\begin{aligned}&
\begin{aligned}
i\hbar \dot{c}_0 &=\frac{\partial\langle \hat H \rangle}{\partial c_0^*} \\
&=n U_\Lambda c_0+2\frac{U_\Lambda}{V}  \sum_{\vec k\cdot \hat{\vec z}>0}^\Lambda\frac{c_0^* g_{\vec k}+c_0\left|g_{\vec k}\right|^2}{1-\left|g_{\vec k}\right|^2}
\end{aligned}\\&
\begin{aligned}
i\hbar \dot{g}_{\vec k} &= \left(1-\left|g_{\vec k}  \right|^2  \right)\frac{\partial \langle \hat H \rangle}{\partial g_{\vec k}^*}
\\& =2\left(\epsilon_{\vec k}+nU_\Lambda  \right)g_{\vec k}+\frac{U_\Lambda}{V}\left[c_0^2+c_0^{*2}g_{\vec k}^2+2\left|c_0  \right|^2g_{\vec k}  \right] \\&
\quad\quad +2\frac{U_\Lambda}{V}\sum_{\vec q\cdot\hat{\vec z}>0}^\Lambda \frac{2\left|g_{\vec q}  \right|^2g_{\vec k}+g_{\vec q}+g_{\vec q}^*g_{\vec k}^2}{1-\left|g_{\vec q}  \right|^2}
\end{aligned}
\end{aligned}.
\end{equation}
Assuming spherical symmetry for $g_{\vec k}$, we integrate these coupled differential equations numerically. The short-time dynamics are essentially cutoff independent for length scales  $r\gg\Lambda^{-1}$ as long as $\Lambda$ is chosen to be much larger than the other momentum scales of the problem, such as $n^{1/3}$ or $a_f^{-1}$. As discussed in Ref. \cite{Sykes2014}, we could equivalently simulate the short-range interactions with attractive square or Gaussian wells of range $r_0$, whose depths are tuned to give the correct scattering length $a_f$. The results for length scales $r\gg r_0$ are then independent of $r_0$ as long as the gas is dilute $n r_0^3\ll 1$.

It is interesting to note that our equations of motion (derived from an ansatz for the quantum state) map directly onto the time-dependent Hartree-Fock-Bogoliubov (HFB) formulation. In that case, one expands the Heisenberg-picture field operator as
\begin{equation}
\hat \psi (\vec r,t) \approx \Phi_0(t) + \frac{1}{\sqrt{V}}\sum_{\vec k\neq 0} e^{i\vec k\cdot \vec r}\left(u_{\vec k}(t)\hat{b}_{\vec k}+v^*_{\vec k}(t)\hat{b}_{-\vec k}^\dagger  \right),
\end{equation}
where $\Phi_0$ is the condensate component and the $u_{\vec k}(t)$ and $v_{\vec k}(t)$ are quasiparticle amplitudes. After making certain mean-field approximations \cite{Griffin1996}, one can write coupled equations of motion for $\Phi_0(t)$ and $\{u_{\vec k}(t),v_{\vec k}(t)  \}$. It is then straightforward to show that our parameter $g_{\vec k}(t)$ and the HFB quantity $v_{\vec k}^*(t)/u_{\vec k}^*(t)$ satisfy exactly the same equations of motion. This equivalence between our variational calculation and the HFB formalism was suggested recently in Ref. \cite{Rancon2014}. Our variational treatment thus suffers from the same low-momentum energy gap as found in HFB; however, because this unphysical gap should manifest itself at longer time and length scales, this should not hinder our study of short-time, short-distance behavior. This is a motivating reason why HFB was able to correctly simulate \cite{Kokkelmans2002} the coherent atom-molecule oscillations observed a decade ago \cite{Donley2002}, the main results of which can be reproduced by our single-channel variational model \cite{footnote1}.

\begin{figure}
\includegraphics[width=0.45\textwidth]{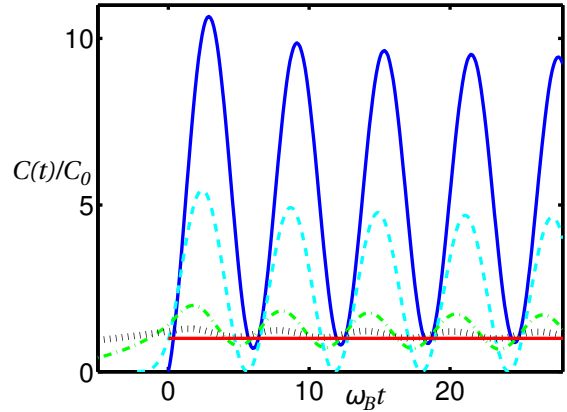}
\caption{(Color online) 
Contact dynamics following a quench from noninteracting to $700 a_0$, for several ramp speeds near the $^{85}$Rb Fano-Feshbach resonance at $155.04$ G. We assume a density of $10^{12}~\mathrm{cm}^{-3}$. As a reference, the red (horizontal) line represents the prediction from unregularized Bogoliubov theory, $C_0=16\pi^2n^2 a_f^2$, which specifies the units of the plot. The blue (oscillating, solid) line is the many-body-variational prediction for a quench that is completely diabatic, the cyan (dashed) line is for the experimental ramp speed of $ \dot{B}_{\mathrm{exp}}~=~1.6~\mathrm{G} / \mu s$ \cite{Makotyn2014}, the green (dot-dashed) line is for a ramp speed of $\dot{B}_{\mathrm{exp}}/10$, and the black (dotted) line is for a ramp speed of $\dot{B}_{\mathrm{exp}}/50$. In each case, the time $t=0$ defines the end of the magnetic field ramp.
 }
\label{fig: adiabatic quench}
\end{figure}

In our accounting of finite ramp speeds, we allow the scattering length to be time dependent in the Hamiltonian, Eq.~\eqref{eq: hamiltonian}. Without loss of generality, we model these ramps using the scattering length profile of the $^{85}$Rb Fano-Feshbach resonance at $B_{0}=155.04~\mathrm{G}$:
\begin{equation}
a_s(t) = a_{\mathrm{bg}}\left(1-\frac{\Delta}{B(t) - B_0}  \right),
\end{equation}
where the resonance width is $\Delta = 10.7~\mathrm{G}$ and the background scattering length is $a_{\mathrm{bg}}=-443~a_0$ \cite{Claussen2003}. We assume linear ramps in the magnetic field. At all instants in time, we find that there is a well-defined $k^{-4}$ tail, which we use to extract the dynamical contact. Figure \ref{fig: adiabatic quench} shows the time evolution of the contact for a noninteracting BEC of density $n = 10^{12}~\mathrm{cm}^{-3}$ that is quenched at several speeds to a final scattering length of $a_f=700~a_0$. We see that the contact oscillates at approximately the frequency of the bound state, $\omega_B$, and the contrast of these oscillations is strikingly large even when we account for the finite experimental ramp rate of $\dot{B}_{\mathrm{exp}}=1.6~\mathrm{G} / \mu s$ reported in Ref.~\cite{Makotyn2014}. The nature of the interference leading to these oscillations will become apparent in the careful two-body calculation of Sec.~\ref{sec: two body}. We stress that these dynamics are quite different from those predicted by unregularized Bogoliubov theory.

 For the case of an instantaneous quench, it is difficult to define a contact in the unregularized theory because the momentum distribution does not have a well-defined $k^{-4}$ tail \cite{Natu2013}. However, such a tail exists as long as the ramp time $t_R$ is nonzero, and it occurs at momenta such that $\hbar k^2 t_R / m \gg 1$. Large-momentum quasiparticles adiabatically follow the scattering length in this case, and the contact thus saturates quickly to a new equilibrium value $C_0 = 16\pi^2 n^2 a_f^2$ over the arbitrarily-small time scale of the quench, regardless of the initial scattering length. These trivial dynamics are plotted as the horizontal red line in Fig.~\ref{fig: adiabatic quench}, and they are in stark contrast to the strong oscillatory behavior predicted by the regularized theory. The peak-to-trough oscillation amplitude remains as large as $C_0$ itself when the experimental ramp speed is decreased by a factor of $10$. Further decreasing the ramp rate eventually results in a quench that is adiabatic with respect to the bound state, in which case the regularized and unregularized theories agree and give a nonoscillating contact.

One can sense the limitations of the unregularized Bogoliubov description of diabatic quench experiments by considering momentum cutoffs. If one uses a coupling constant $U_\Lambda \rightarrow 4\pi\hbar^2 a_f/m$ in the Hamiltonian given by Eq.~\eqref{eq: hamiltonian}, as is typically required in mean-field theories at the Bogoliubov level, then we see from Eq.~\eqref{eq: coupling constant} that this implies a momentum cutoff $\Lambda$ satisfying $\Lambda a_f \ll 1$ \cite{Braaten1997,Pethick}. There is no bound state in this limit, and any important physics occuring uniquely on the time scale $\omega_B^{-1}$ and length scale $a_f$ of the bound state is therefore absent in all variants of unregularized Bogoliubov theory \cite{Natu2013,Yin2013,Kain2014,Rancon2014}. In the limit that the quench is adiabatic with respect to the bound state, the time scale of the quench is at least consistent with an implied energy cutoff  $\hbar^2\Lambda^2/m\ll \hbar\omega_B$. That is precisely the regime in which the unregularized Bogoliubov theory correctly describes the contact dynamics, as shown in Fig. \ref{fig: adiabatic quench}.

\begin{figure}
\includegraphics[width=0.45\textwidth]{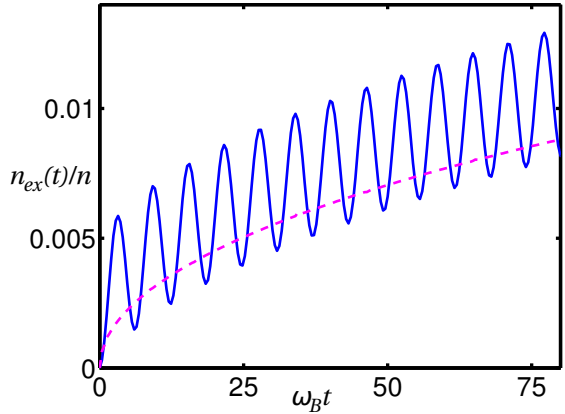}
\caption{(Color online) 
Excitation fraction following a diabatic quench from noninteracting to $700 a_0$, for a BEC of density $n=10^{12}~\mathrm{cm}^{-3}$. The blue (solid) line is the prediction of our many-body-variational formulation. The magenta (dotted) line is the prediction from unregularized Bogoliubov theory.
 }
\label{fig: depletion}
\end{figure}

These bound-state oscillations also occur in the depletion of the condensate. Figure \ref{fig: depletion} shows the time evolution of the depletion fraction $n_{\rm{ex}}/n$ after a diabatic quench, computed both with the regularized and unregularized theories. In the unregularized theory (cf. Ref. \cite{Natu2013}), the depletion grows as $\sqrt{t}$, which captures the overall growth of the depletion except for the large-amplitude oscillations. Similar oscillations have been discussed previously in the theory literature \cite{Timmermans1999,Holland2001,Duine2003a,Duine2004}, and they received only a brief mention in the experimental results of Ref. \cite{Donley2002}. To date, there is no published data on the subject, although some preliminary observations of these condensate oscillations can be found in Ref. \cite{ClaussenThesis}.

\section{Two-Body Models} \label{sec: two body}

Two-body models afford another intuitive description of BEC quench dynamics. In some cases, they are exactly solvable \cite{Busch1998}, and they can paint relatively-transparent pictures of nonequilibrium physics \cite{Boschi2014,Goral2005,Mies2000,Borca2003, Goral2004} that are sometimes obscured by the mathematics of more sophisticated, many-body models. It was recently suggested that such models might even be made quantitatively accurate in their descriptions of short-time, large-momentum dynamics in quenched BEC systems \cite{Sykes2014}, although an unambiguous, universal link to the many-body BEC problem has been absent in the literature. In this section, we establish such a link and derive analytic predictions for the contact dynamics following a diabatic quench of the scattering length near a broad Fano-Feshbach resonance.

\subsection{Calibration}

We consider the quantum dynamics of a pair of free-space atoms whose scattering length is quenched from an initial value $a_i\geq 0$ to some finite value $a_f\geq 0$. Assuming zero momentum for the center of mass (as is the case for any pair of atoms that scatter out of a BEC), the post-quench dynamics are governed by the time-dependent Schr\"{o}dinger equation
\begin{equation} \label{eq: two body hamiltonian}
i\hbar \frac{\partial \psi(\vec r,t)}{\partial t}= -\frac{\hbar^2}{2\mu}\nabla^2\psi(\vec r,t) +\frac{2\pi\hbar^2 a_f}{\mu}\delta(\vec r)\frac{\partial}{\partial r}\left[r\psi(\vec r,t)  \right]
\end{equation}
where $\psi(\vec r,t)$ is the wavefunction for the relative coordinate $\vec r$, $\mu=m/2$ is the reduced mass, and we model the short-range interactions of the system with the Fermi pseudopotential \cite{Fermi1936,Huang1957}. This pseudopotential is equivalent to the $\Lambda\rightarrow \infty$ limit of the regularized contact interaction given by Eqs. \eqref{eq: hamiltonian}-\eqref{eq: coupling constant}. 

We can time-evolve an arbitrary spherically-symmetric initial condition by expanding in the basis of energy-normalized $s$-wave eigenfunctions of the Hamiltonian shown in Eq.~\eqref{eq: two body hamiltonian}. These eigenfunctions are 
\begin{equation} \label{eq: scattering states}
\psi_k^{(S)}(r)=\frac{\mathrm{sin}(kr)-k a_f \mathrm{cos}(kr)}{r\sqrt{4\pi^2\hbar^2k(k^2a_f^2+1)/m}} \quad,\quad E_k=\frac{\hbar^2k^2}{m}
\end{equation}
for the scattering states and 
\begin{equation} \label{eq: bound state}
\psi_B(r)=\frac{\mathrm{e}^{-r/a_f}}{r\sqrt{2\pi a_f}}\quad,\quad E_B=-\frac{\hbar^2}{m a_f^2}
\end{equation}
for the bound state. Given an initial condition $\psi_0(r)$, the solution to Eq.~\eqref{eq: two body hamiltonian} is \cite{Merzbacher}
\begin{equation}\label{eq: position space solution}
\begin{aligned}
\psi(r,t)= & \int\limits_0^\infty dE_{k'} \mathrm{e}^{-i E_{k'} t}\psi_{k'}^{(S)}(r)\int d^3r' \psi_{k'}^{(S)}(r') \psi_0(r') \\&
\quad + \mathrm{e}^{-i E_B t}\psi_B(r) \int d^3r'\psi_B(r')\psi_0(r)
\end{aligned}.
\end{equation}
We then evaluate the momentum distribution by taking the Fourier Transform of Eq.~\eqref{eq: position space solution}:
\begin{equation} \label{eq: f transform}
\psi(\vec k,t)=\int d^3r \mathrm{e}^{-i \vec k \cdot \vec r}\psi(r,t) .
\end{equation}

Following our intuition from mean-field theory, we relate this two-body problem to the many-body system by considering the combined effect of a background, dilute BEC on the momentum distribution of a single particle. Assuming that this time-dependent, single-particle momentum distribution is normalized in the continuum via
\begin{equation}
1=\int \frac{d^3k}{(2\pi)^3}\left|\psi(\vec k,t)\right|^2,
\end{equation}
we compute the full momentum distribution by multiplying by the total density $n$ \cite{Mies2000,Goral2005}. The combined effect of the dilute background gas is modeled by an appropriate choice of initial condition $\psi_0(r)$. Previous calculations of this type have placed the two-body system in a (fictitious) tight harmonic trap, whose frequency is chosen to reproduce either the total density $n$ \cite{Borca2003} or the approximate mean interparticle separation $\langle r \rangle$ \cite{Sykes2014}. Both of these prescriptions are limited in the sense that their quantitative predictions for short-distance dynamics depend strongly on the harmonic nature of the fictitious trap. In this sense, they are intrinsically semi-quantitative \cite{footnote2}.

The new prescription that we propose is motivated by the fact that a quench of zero-range interactions signifies a quench of a log-derivative boundary condition at $r= 0$:
\begin{equation} \label{eq: log derivative}
\lim_{r\rightarrow 0} \frac{\partial_r \left(r \psi(r)  \right)}{\left(r\psi(r)  \right)}= -\frac{1}{a} .
\end{equation}
As a result, the contact dynamics immediately following a quench occurs entirely in the short range. The most important feature of an initial condition $\psi_0(r)$ is therefore its behavior as $r\rightarrow 0$. Our first requirement is that $\psi_0(r)$ satisfy Eq.~\eqref{eq: log derivative} for the initial scattering length of the system, $a_i$. (All of the eigenstates in our post-quench expansion basis, Eqs.~\eqref{eq: scattering states}-\eqref{eq: bound state}, satisfy this log-derivative condition for the final scattering length, $a_f$.) Importantly, this log-derivative condition does not fix the absolute magnitude of $\psi_0(r)$ for small $r$; any such scaling cancels in Eq.~\eqref{eq: log derivative}. We propose that this absolute scaling of the short-range wavefunction be fixed by the many-body problem. The quantity $\left|\psi_0(r)  \right|^2$ represents the probability density of finding a background particle a distance $r$ from the particle of interest, and this is given by $n g^{(2)}(r)$ in the many body-problem, where $g^{(2)}(r)$ is the two-body correlation function \cite{Pathria}. For a pure, noninteracting BEC, there are no correlations and $g^{(2)}(r) = 1$. If the initial scattering length is nonzero, however, short-range correlations are determined exlusively by the contact via Eq.~\eqref{eq: g2}. We thus calibrate the short-range behavior of the two-body wavefunction as follows:
\begin{subequations} \label{eq: calibrations}
\begin{equation} \label{eq: calibration 1}
\lim_{r\rightarrow 0} \left|\psi_0(r)  \right|^2= n \quad\quad\quad\quad\quad\quad \quad\quad (a_i=0)
\end{equation}
\begin{equation} \label{eq: calibration 2}
 \left|\psi_0(r)  \right|^2 \rightarrow \frac{C_i}{16\pi^2 n r^2}+\mathcal{O}\left(\frac{1}{r}  \right)\quad\quad (a_i>0)
\end{equation}
\end{subequations}
where $C_i=16\pi^2 n^2 a_i^2$ is the contact for the initial, dilute BEC \cite{Lee1957,Wild2012}. Equation \eqref{eq: calibration 2} guarantees that the contact for the particle of interest (within our simple model) matches the particle-number-averaged contact of the many-body system \cite{footnote3}.

We now choose a set of initial two-body wavefunctions to test the robustness of our calibration scheme. We have been able to analytically evaluate the integrals in Eqs.~\eqref{eq: position space solution}-\eqref{eq: f transform} for the initial conditions
\begin{equation} \label{eq: initial wave functions}
\psi_0(r)= 
\begin{cases}
A_0(a_i,L_0)\left(1-\frac{L_0 a_i}{(L_0-a_i)r}  \right) \mathrm{e}^{-r/L_0} \\
A_1(a_i,L_1)\left(1-\frac{a_i}{r}  \right)\mathrm{e}^{-r/L_1}\left[1+\frac{r}{L_1}  \right] \\
A_2(a_i,L_2)\left(1- \frac{a_i}{r} \right)\mathrm{e}^{-r/L_2}\left[1+\frac{r}{L_2}+\frac{1}{2}\left(\frac{r}{L_2}  \right)^2  \right]  
\end{cases}
\end{equation}
where $L_j$ is a free parameter for each initial condition, and $A_j(a_i,L_j)$ is a normalization constant. The leading factor in parentheses enforces the log-derivative boundary condition, and the bracketed polynomial factors have been chosen to add variety to our treatment of the long-range wavefunction. Recall that the calibration given by Eq.~\eqref{eq: calibrations} completely specifies the short-range behavior, along with the free parameter $L_j$. The necessary integrations in Eqs.~\eqref{eq: position space solution}-\eqref{eq: f transform} can be carried out with a combination of contour integration and symbolic mathematical software, along with the useful relation
\begin{equation}
\int\limits_0^\infty dr \mathrm{sin}(k'r)\mathrm{cos}(kr) \rightarrow \mathcal{P}\frac{k'}{k'^2-k^2}
\end{equation}
when integrated against well-behaved functions, with $\mathcal{P}$ denoting the Cauchy Principal Value.

\begin{figure}
\includegraphics[width=0.45\textwidth]{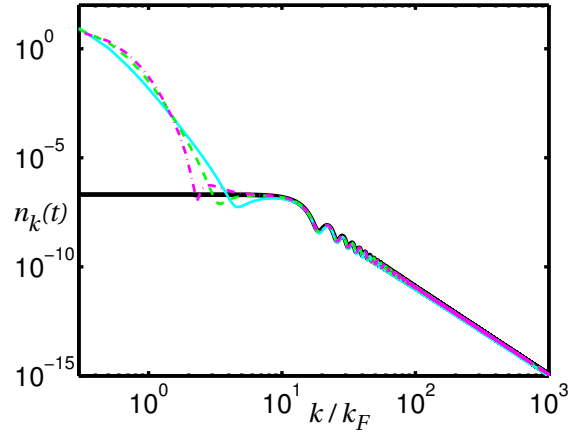}
\caption{(Color online) 
Momentum distributions at fixed time $\omega_F t=0.01$ after a quench from noninteracting to unitarity. The thick black line is the numerical data from a many-body variational calculation. The thin lines are the analytically-computed two-body results. The cyan (solid), green (dashed), and magenta (dot-dashed) lines respectively correspond to the properly-calibrated initial conditions in the order listed in Eq.~\eqref{eq: initial wave functions}.
 }
\label{fig: momentum distributions}
\end{figure}

A useful figure of merit for short-range nonequilibrium physics is the slope of the contact growth after a quench from noninteracting ($a_i=0$) to unitarity ($a_f=\infty$). Using the many-body variational formulation described in Sec.~\ref{sec: mean field}, Ref.~\cite{Sykes2014} found previously that the contact grows as $C(t)\approx 26.9 n^{4/3} \omega_F t$ at short times, where $\omega_F = \hbar k_F^2 / 2m$ is the Fermi frequency of the gas and $k_F=\left(6\pi^2 n  \right)^{1/3}$ is the Fermi momentum. For our two-body models, the formulae for the exact momentum-space wave functions $\psi(k,t)$ are too lengthy to reproduce here (see Appendix \ref{app 1} for an example); however, their predicted momentum distributions are plotted in Fig. \ref{fig: momentum distributions} at a fixed time shortly after the quench, and they are compared with the many-body prediction. With the free parameter of each initial condition chosen in our prescribed manner, all results agree favorably at large momentum. Despite the various functional forms for the initial conditions in Eq.~\eqref{eq: initial wave functions}, all of the two-body wavefunctions predict that
\begin{equation} \label{eq: linear contact}
C(t) = \frac{128 \pi}{\left( 6\pi^2\right)^{2/3}} n^{4/3} \omega_F t,
\end{equation}
at short times, which agrees with the many-body variational prediction to within less than two percent. Equation \eqref{eq: linear contact} also follows from applying our prescription to the Gaussian initial condition of Ref. \cite{Sykes2014}, for which only the contact growth can be calculated analytically. The contact slope now appears to be independent of the arbitrary details of the two-body model.

We remark that the model independence of our large-momentum dynamics is nontrivial. If, instead, we choose each free parameter $L_j$ by matching $\langle r \rangle$ to the nearest-neighbor separation, the predicted slope of the contact varies by almost an order of magnitude, depending on the chosen initial condition. The approximate agreement between the two- and many-body models demonstrated in Ref. \cite{Sykes2014} is a result of the near-equivalence of the requirements that $\langle r \rangle\equiv \left(4 \pi n /3  \right)^{-1/3}$ and $\left|\psi_0(0)  \right|^2\equiv n$ for normalized Gaussian functions. As explained above, the latter requirement is more physically motivated, and it leads to improved agreement with the many-body results while unifying the large-momentum predictions of the various exactly-solvable two-body models. In the remainder of our discussion, we employ this calibration scheme.

Contact dynamics aside, our models also agree on the subleading oscillatory structure of the large-momentum dynamics, as shown in Fig.~\ref{fig: momentum distributions}. These oscillations have phase $E_k t /\hbar$, and their amplitude scales as $k^{-5}$, as found previously \cite{Sykes2014}. Each distribution shows distinct low-momentum behavior that is determined by the long-range characteristics of the initial conditions. We can infer from Eq.~\eqref{eq: initial wave functions} that these long-range features occur on a length scale that is set by the parameter $L_j$, which is of the order of the mean interparticle spacing for the gas. At such momentum scales, we expect many-body effects to determine the physics, and this limits the approximate validity of our two-body models to momenta $k\gg k_F$ and times $\omega_F t \ll 1$.

\subsection{Quenching to Finite Scattering Length}

With our two-body models properly calibrated, we are well equipped to revisit and generalize the bound-state oscillations addressed in Sec.~\ref{sec: mean field}. We will see that the simple two-body approach illustrates the crucial role played by the bound state after a diabatic quench, while quantitatively describing the evolution of two-body correlations via the dynamical contact.

As a preliminary matter, our two-body approach leads to an intuitive understanding of bound-state oscillations. The basic structure of Eq.~\eqref{eq: position space solution} suggests that the bound and scattering states may be compared to the two legs of a simple interferometer. The diabatic quench essentially projects the initial condition onto these two legs, and a different phase is acquired over each leg as time progresses, as evidenced by Eq.~\eqref{eq: position space solution}. The measured momentum distribution is always defined with respect to free-particle (noninteracting) momentum states, rather than the scattering states of Eq.~\eqref{eq: scattering states}; it is for this definition that the $k^{-4}$ tail is meaningfully related to short-range density-density correlations via the contact \cite{Tan2008a,Stewart2010}. Hence, the two legs of the interferometer are recombined during a measurement of the momentum distribution, thereby projecting the quantum state onto the free-particle momentum basis as in Eq.~\eqref{eq: f transform}. The phase evolution of the bound-state component leads to periodically-modulated interference that is most pronounced at the length scale of the bound state, $r\lesssim a_f$. As a result, the contact oscillates, along with certain other observables such as the condensate fraction (see Fig.~\ref{fig: depletion}).

\begin{figure}
\includegraphics[width=0.45\textwidth]{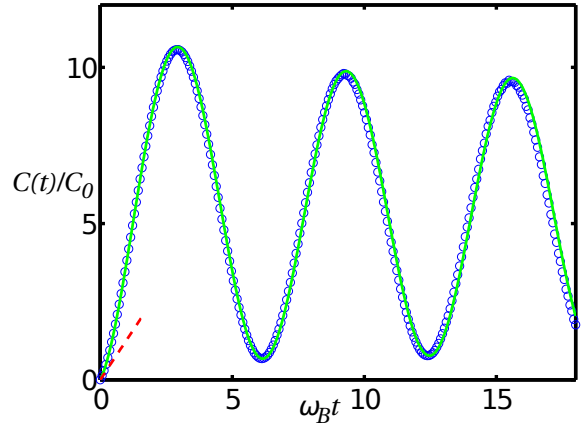}
\caption{(Color online) 
Contact dynamics following a diabatic quench from noninteracting to $700~a_0$, for a BEC of density $n=10^{12}~\mathrm{cm}^{-3}$. The circles represent the many-body-variational data (shown also in Fig. \ref{fig: adiabatic quench}). The green (solid) line represents the formula given in Eq.~\eqref{eq: contact dynamics}, and the red (dotted) line represents the linear growth given in Eq.~\eqref{eq: linear contact}.
 }
\label{fig: few and many}
\end{figure}

As discussed previously, the various initial conditions of Eq.~\eqref{eq: initial wave functions} lend themselves to analytical, time-dependent solutions for arbitrary initial and final scattering lengths. These formulae are quite complicated in general, but, remarkably, they each predict the same behavior of the contact \begin{equation} \label{eq: contact dynamics}
\begin{aligned}
C(t)=16 & \pi^2 n^2 a_f^2  \\&
 \times \left| 1+\left(\frac{a_i}{a_f}-1  \right)\mathrm{e}^{i \omega_B t}\left( 1+\mathrm{erf}\left[\sqrt{i \omega_B t}  \right] \right)  \right|^2 
\end{aligned}.
\end{equation}
if we take the limits $na_i^3 \ll 1$ and $n a_f^3 \ll 1$. Figure~\ref{fig: few and many} plots Eq.~\eqref{eq: contact dynamics} against the many-body data for the diabatic quench considered already in Fig.~\ref{fig: adiabatic quench}. Apart from a slight offset in the oscillation frequency, the agreement is excellent. We believe that this small frequency deviation is due to the fact that our numerical solution of the many-body model is constrained to a finite (albeit large) momentum cutoff $\Lambda$, whereas our two-body models are truly zero-range. Any experimental realization of these oscillations would experience such an offset due to the finite range of true interatomic interactions. This was certainly the case in the Ramsey experiment of Ref.~\cite{Donley2002}. Aside from the bound-state oscillations of the contact, the momentum distributions look essentially the same as in Fig.~\ref{fig: momentum distributions}, including the subleading $k^{-5}$ behavior mentioned previously.

It is useful to examine the general dynamics given by Eq.~\eqref{eq: contact dynamics}. At short times $\omega_B t \ll 1$, the contact evolves continuously from its intial value $C_i$ as
\begin{equation} \label{eq: contact growth 2}
\begin{aligned}
C(t)=C_i+&32\pi^2n^2 a_i \left(a_i-a_f\right)\sqrt{\frac{2}{\pi}\omega_B t} \\&+\frac{128\pi}{\left(6\pi^2  \right)^{2/3}}\left(\frac{a_i}{a_f}-1\right)^2 n^{4/3} \omega_F t +\mathcal{O}(t^{3/2})
\end{aligned}.
\end{equation}
In the limit of vanishing initial scattering length, the contact first grows linearly according to Eq.~\eqref{eq: linear contact} for all values of $a_f$. This is shown in Fig.~\ref{fig: few and many} for the case of a quench to $a_f=700 a_0$. However, at nonzero initial scattering length, this linear growth is superseded by nonanalytic $\sqrt{t}$ behavior. At later times $\omega_B t \gg 1$, the contact is oscillatory:
\begin{equation} \label{eq: contact oscillations}
\begin{aligned}
C(t)\approx 16\pi^2 n^2 a_f^2 \Bigg[ 1+ & 4 \left( \frac{a_i}{a_f}-1 \right)^2  \\& 
 \quad+ 4\left(\frac{a_i}{a_f}-1  \right)\mathrm{cos}\left(\omega_B t \right)\Bigg]
\end{aligned} .
\end{equation}
For a diabatic quench upward ($a_f>a_i$), the time-averaged contact $\langle C(t) \rangle_t$ may be up to five times larger than the unregularized Bogoliubov prediction of $C_0=16\pi^2n^2 a_f^2$, and the oscillation amplitude may be up to four times as large. Of course, in the limit of no quench ($a_f= a_i$), the contact is trivially time independent \cite{footnote4,footnote5}.

\begin{figure}
\includegraphics[width=0.45\textwidth]{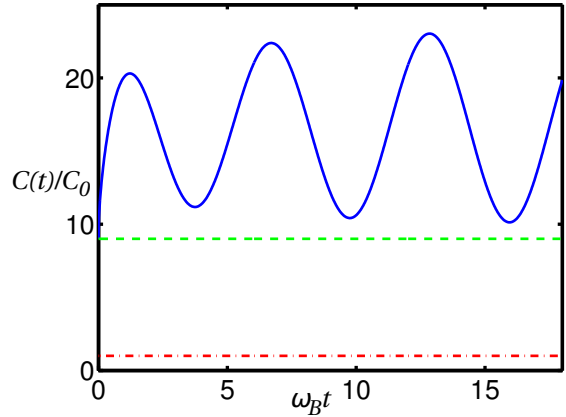}
\caption{(Color online) 
Contact dynamics following a diabatic quench downwards with $a_f = a_i/3$. For reference, the green (dotted) line is the equilibrium contact $C_i$ for the BEC at the initial scattering length $a_i$, and the red (dot-dashed) line is the contact $C_0$ for a ground state BEC at the final scattering length $a_f$.
 }
\label{fig: quench down}
\end{figure}

The case of a diabatic quench downward ($a_f < a_i$) reveals interesting physics. Depending on the ratio of initial and final scattering lengths, the time-averaged contact may be much larger than the unregularized Bogoliubov prediction $C_0$, and larger even than the initial contact $C_i$. Figure \ref{fig: quench down} shows these dynamics for a quench to $a_f = a_i / 3$, in which case $\langle C(t) \rangle_t$ is more than an order of magnitude larger than $C_0$ and almost twice as large as $C_i$. The peak-to-trough oscillation amplitude is also larger than both $C_0$ and $C_i$. This is in stark contrast to the unregularized Bogoliubov case, in which the contact relaxes to $C_0$ over the fast timescale of the diabatic quench. At least in the transient dynamics, a diabatic reduction in the scattering length can evidently increase local two-body correlations by up to a factor of four compared to the initial condition, as evidenced by Eq.~\eqref{eq: contact oscillations}. An important limitation, however, is that these dynamical correlations are most pronounced at and below the length scale of the bound state.

The heightened short-range correlations contained in $\langle C(t) \rangle_t$, beyond those already at the unregularized Bogoliubov level, come fundamentally from bound-state physics. For example, if we ignore the contribution of the bound state to the dynamics in Eq.~\eqref{eq: position space solution}, we find that the scattering states dephase in such a way that the dynamical contact asymptotes towards $C_0$ without any oscillations, regardless of the initial scattering length. This is in qualitative agreement with the unregularized Bogoliubov prediction. Once the scattering states dephase, however, the bound state is left to dominate the short-range wavefunction except in the relatively trivial case where $a_f \sim a_i$. The excess short-range correlations, given by the second bracketed term in Eq.~\eqref{eq: contact oscillations}, are determined entirely by the original projection of the initial condition onto the bound state.

We can estimate the amplitude of the depletion oscillations by simply integrating over the relevant portion of the momentum distribution. From Eq.~\eqref{eq: contact oscillations} and the fact that $n_k(t)\sim C(t) / k^4$ at large $k$, the part of the momentum distribution oscillating at the bound-state frequency behaves as
\begin{equation} \label{eq: momentum oscillations}
n_{k,\mathrm{osc}}(t)\sim \frac{16\pi^2n^2 a_f^2}{k^4}4\left(\frac{a_i}{a_f}-1  \right)\mathrm{cos}\left(\omega_B t  \right) ,
\end{equation}
aside from the time-independent contribution to the $k^{-4}$ tail. As mentioned previously, these oscillations occur at the momentum scale of the bound state, where $k a_f \gtrsim 1$. Integrating Eq.~\eqref{eq: momentum oscillations}, we find that the oscillating part of the depletion fraction is approximately
\begin{equation}
\begin{aligned}
\frac{n_{\mathrm{ex , osc}}(t)}{n} & \sim \frac{1}{n} \int\limits_{k a_f >1} \frac{d^3 k}{(2\pi)^3} n_{k, \mathrm{osc}}(t) \\&
\sim 32 \left( n a_f^3 \right)\left(\frac{a_i}{a_f}-1  \right)\mathrm{cos}\left( \omega_B t \right)
\end{aligned}.
\end{equation}
The oscillation amplitude given here agrees with the many-body data shown in Fig.~\ref{fig: depletion} to within a factor of order unity, and we expect it to be a reasonable estimate as long as the diluteness parameter $n a^3$ is small before and after the diabatic quench.

As a final aside, we note that our two-body analysis is able to generalize the short-time dynamics following a quench to unitarity. Although we have presented the derivation of Eq.~\eqref{eq: contact growth 2} for the case of weak interactions ($n a_i^3\ll 1$, $n a_f^3 \ll 1$), we have found that its $a_f\rightarrow \infty$ limit is well defined and correctly describes the short-time contact dynamics after diabatically quenching the two-body wavefunction from weakly interacting to unitarity:
\begin{equation} \label{eq: unitarity growth}
C(t)=C_i-32\pi^2n^2 a_i\sqrt{\frac{2\hbar}{\pi m} t} +\frac{128\pi}{\left(6\pi^2  \right)^{2/3}} n^{4/3} \omega_F t .
\end{equation}
This limit represents a generalization of Eq.~\eqref{eq: linear contact} for diabatic quenches from small initial scattering length $a_i \geq 0$, and it agrees with the earlier result when $na_i^3$ is vanishingly small.

\section{Conclusion} \label{sec: conclusion}

We have elucidated the important role of the bound state in determining the contact dynamics of a diabatically-quenched BEC. We first computed these dynamics using a variational many-body model, demonstrating that large-amplitude oscillations of the contact can be observed even with existing magnetic-field-ramp technology. Our calculations reinforce the idea that coherent, short-range physics can lead to measureable signatures even in the BEC fraction. This is the dominant physics of the quenched gas on short time scales, before many-body effects and loss become important.

We also developed a calibration scheme for two-body models that leads to an unambiguous, quantitative description of BEC contact dynamics following a sudden quench. Our prescription fixes both the log-derivative and absolute magnitude of the initial short-range, two-body wavefunction by matching to the many-body problem, and we are able to derive analytic formulae for the short-time evolution of the contact in the weakly-interacting and unitarity limits. Our computed dynamics are shown to be independent of the arbitrary features of the models, and they agree with many-body predictions. This two-body picture indicates that bound-state oscillations of the contact are analogous to interferometry. We expect that one can account for finite ramp speeds by numerically solving the two-body Schrodinger equation for a properly-calibrated model \cite{footnote4}.

The dynamical contact can be measured using time-resolved RF spectroscopy, as done in Ref.~\cite{Bardon2014}. Our analysis shows that even the time-averaged contact $\langle C(t) \rangle_t$ may be greatly magnified relative to the unregularized Bogoliubov prediction due to bound-state physics, and this could be observed with an RF pulse that is long compared to the bound-state oscillation period. Measuring the oscillations themselves necessarily requires using shorter pulses, and that may lead to inconvenient broadening of the central RF peak. In any event, the temporal constraints on time-resolved RF spectroscopy depend both on the atomic species and on the transition under consideration, and they are beyond the scope of the present study.

We reiterate that the bound-state dynamics that we have considered are a coherent, transient effect. They encapsulate the response of a short-range wavefunction to an abrupt change in the scattering length or, equivalently, a log-derivative boundary condition. At longer time scales, we expect the oscillations to damp out as the system equilibrates. Similar damping was observed in the Ramsey experiment of Refs.~\cite{Donley2002,Claussen2003}, and it was believed to be due to a combination of incoherent three-body loss and dephasing from magnetic-field inhomogeneities. Still, the coherence of large-momentum dynamics persisted for many oscillation periods before damping became significant. The engineering of quench apparatus has improved over the years, especially in creating ramps that are diabatic with respect to the bound state \cite{Makotyn2014}. This opens the door for systematic experimental studies of bound-state signatures in quenched BECs.

\section*{Acknowledgement}
J.P.C and J.L.B recognize support from the NDSEG fellowship program and the NSF, respectively. We acknowledge helpful conversations from A. G. Sykes and K. R. A. Hazzard.

\appendix

\section{Sample Momentum Distribution} \label{app 1}
The time-dependent, two-body momentum distribution can be calculated analytically for each of the initial conditions listed in Eq.~\eqref{eq: initial wave functions}, and for arbitrary initial and final scattering lengths. Except in certain limits (discussed in the main text), these formulas are generally too lengthy to usefully write down. We include here the simplest example, which is the distribution for the bare exponential of Eq.~\eqref{eq: initial wave functions} after a quench from noninteracting ($a_i=0$) to unitarity ($a_f= \infty$). After evaluating the integrals in Eqs.~\eqref{eq: position space solution}-\eqref{eq: f transform}, we find that
\begin{equation} \label{eq: wave function formula}
\begin{aligned}
\psi(k,t)= &\frac{8\sqrt{L_0}}{k\left(1+k^2L_0^2\right)^2} \times \\& \quad
\left\{\left(1-k^2L_0^2  \right)\mathrm{DawsonF}\left[\sqrt{i\frac{\hbar k^2 t}{m}}  \right]\right. \\& \quad\quad\left.
+\left(1+k^2L_0^2 \right)\sqrt{i\frac{\hbar k^2 t}{m}}\right. \\&\quad\quad
\left.  +\mathrm{e}^{i\frac{\hbar t}{ m L_0^2}}\sqrt{\pi}\left(kL_0-i\left(1+k^2L_0^2\right)\frac{\hbar k t}{mL_0}  \right)\right. \\&\quad\quad\quad\quad\quad \left.
\times \mathrm{erfc}\left[\sqrt{i\frac{\hbar t}{mL_0^2}}  \right]  \right\}
\end{aligned} ,
\end{equation}
where the Dawson function is defined by
\begin{equation}
\mathrm{DawsonF}(z)\equiv \mathrm{e}^{-z^2}\int_0^z dy ~\mathrm{e}^{y^2} .
\end{equation}
This wavefunction evolves continuously from its initial condition.


\begin{thebibliography}{99}
\bibitem{footnote0} On a theoretical level, one can also countenance diabatic quenches of the scattering length for purely repulsive interaction potentials, which have no bound state. The physics is certainly different in such cases. However, the $a_s>0$ side of a Fano-Feshbach resonance always admits a bound state.
\bibitem{footnote1} Our single-channel model can reproduce the correct molecular binding energy only in the regime where the bound state displays universal features, such as $E_B~\rightarrow~-~\hbar^2/ma_s^2$. To recover the species-specific features of the bound state, one must use a multichannel model.
\bibitem{footnote4} In our two-body models, the contact dynamics for the case of $a_f=a_i$ are nontrivial at higher order in $na^3$. One can correct for this by choosing an initial condition that, besides being calibrated according to Eqs.~\eqref{eq: log derivative}-\eqref{eq: calibration 2}, is orthogonal to the initial bound state at scattering length $a_i$. Such an initial condition would be consistent with that of a metastable interacting BEC. This may be important to future two-body calculations, for instance, when taking account of the nondiabaticity of slower magnetic-field ramps or when higher-order corrections in $na^3$ are required.
\end{thebibliography}
\end{document}